\documentclass[aps,floats,floatfix,superscriptaddress]{revtex4}
\usepackage{amsfonts,amsmath,amssymb,hyperref}
\usepackage{filecontents}
\usepackage[pdftex]{graphicx}

\usepackage{color}
\usepackage{times}

\usepackage{multirow}

\renewcommand{\geq}{\geqslant}

\begin{document}

\title{Reporting delays: a widely neglected impact factor in COVID-19 forecasts}
\author{Long Ma}
\affiliation{Faculty of Electrical Engineering, Mathematics, and Computer Science, Delft University of Technology, 2600 GA Delft, The Netherlands}
\author{ Piet Van Mieghem}
\affiliation{Faculty of Electrical Engineering, Mathematics, and Computer Science, Delft University of Technology, 2600 GA Delft, The Netherlands}
\author{Maksim Kitsak}
\affiliation{Faculty of Electrical Engineering, Mathematics, and Computer Science, Delft University of Technology, 2600 GA Delft, The Netherlands}

\begin{abstract}

Epidemic forecasts are only as good as the accuracy of epidemic measurements. Is epidemic data, particularly COVID-19 epidemic data, clean and devoid of noise? Common sense implies the negative answer. While we cannot evaluate the cleanliness of the COVID-19 epidemic data in a holistic fashion, we can assess the data for the presence of reporting delays. In our work, through the analysis of the first COVID-19 wave, we find substantial reporting delays in the published epidemic data. Motivated by the desire to enhance epidemic forecasts, we develop a statistical framework to detect, uncover, and remove reporting delays in the infectious, recovered, and deceased epidemic time series. Our framework can uncover and analyze reporting delays in 8 regions significantly affected by the first COVID-19 wave. Further, we demonstrate that removing reporting delays from epidemic data using our statistical framework may decrease the error in epidemic forecasts. While our statistical framework can be used in combination with any epidemic forecast method that intakes infectious, recovered, and deceased data, to make a basic assessment, we employed the classical SIRD epidemic model. Our results indicate that the removal of reporting delays from the epidemic data may decrease the forecast error by up to 50\%. We anticipate that our framework will be indispensable in the analysis of novel COVID-19 strains and other existing or novel infectious diseases.
\end{abstract}

\maketitle
\date{\today}

The COVID-19 pandemic has been crippling the world's health, economies, and quality of life for over three years. While  we are likely
 beyond the peak, we are gradually understanding that COVID-19 is here to stay. Fast mutation rates of the virus and its overwhelming spreading capability make it extremely hard if not impossible to eradicate~\cite{phillips2021coronavirus}. COVID-19 is not the first and almost surely not the last pandemic to hit humanity. Therefore, in order to better prepare and withstand other contagious diseases in the future, we need to extract as many lessons as possible from the COVID-19 pandemic.

Public awareness is, arguably, the first line of defense against any infectious diseases. Efficient collection of epidemic data and accurate epidemic forecasts allow for timely containment of the spread or \emph{flattening of the curve} to win time for the development of pharmaceutical treatment methods. Due to the success of network epidemiology and the broad availability of data, we have made significant advances in epidemic modeling and forecast methods~\cite{pastor2015epidemic,kiss2017mathematics,brauer2017mathematical,massimo2020comparing,prasse2022predicting}. However, the accuracy of epidemic forecasts strongly depends on the accuracy and timeliness of the input data. 

Are epidemic data -- especially in the short-time period after the onset of the epidemic -- devoid of noise? Multiple sources suggest the negative answer. One
problem is delays in data reporting~\cite{pellis2020challenges,larremore2021test}, defined as the time difference 
between the event when the person was affected by the virus and the time this event is accounted for. The reporting delays may occur due to many reasons, including patient hesitancy, medical testing delays, and reporting delays by public health authorities~\cite{marinovic2015quantifying,swaan2018timeliness}. As a result, reporting delays may vary not only across different diseases but also across different regions of interest. The median diagnosis delay for Malaria is, for instance, approximately four days \cite{bastaki2018time}. Similarly, the research on the Middle East Respiratory Syndrome CoronaVirus (MERS-CoV) found a time difference between the symptom onset and confirmation of approximately four days \cite{ahmed2017diagnostic}. Hepatitis A, Measles, and Mumps data are usually reported after eight days \cite{marinovic2015quantifying}. Reporting delays for some diseases are significantly longer: Hepatitis B, Shigellosis, and Salmonella can take on average 2-3 weeks \cite{marinovic2015quantifying}.    Recent COVID-19 studies reveal significant reporting delays of infections in China \cite{sun2020early,linton2020incubation,lauer2020incubation,kraemer2020effect,leung2020first,lin2020conceptual}, Italy \cite{cereda2020early}, Germany \cite{dehning2020inferring,gunther2021nowcasting}, Singapore \cite{tariq2020real}, the USA \cite{harris2020overcoming}, and the UK \cite{pellis2020challenges}.

In our work, we develop a statistical framework to de-noise epidemic data by removing reporting delays. We demonstrate that the removal of reporting delays may significantly improve the accuracy of epidemic forecasts. Our statistical framework can be used in combination with any epidemic forecast method that takes infectious, recovered, and deceased epidemic data as input. Our work is organized as follows. We first present the evidence for the presence of reporting delays in the epidemic data extracted from the first COVID-19 wave. We then proceed to develop a statistical framework to remove reporting delays.  After validating our framework on synthetic data, we move on to remove and analyze reporting delays in 8  hotspots of the COVID-19 pandemic. We conclude our work with a discussion of the impact of delay removal on the accuracy of epidemic forecasts.

\begin{table}
%\aboverulesep=0ex
%\belowrulesep=0ex
\centering
\begin{tabular}{cccccccccccc}
%  \toprule
\hline \hline
  \multicolumn{6}{l|}{Cumulative Quantities} & \multicolumn{6}{l}{Quantity Increments} \\
  \hline
%  \midrule
\multicolumn{6}{l|}{$Y$:~fractions of cases} &\multicolumn{6}{l}{$\Delta Y$:~fractions of new cases} \\
\multicolumn{6}{l|}{$\widetilde{Y}$:~fractions of reported cases} &\multicolumn{6}{l}{$\Delta \widetilde{Y}$:~fractions of reported new cases} \\
\multicolumn{6}{l|}{$\hat{Y}$:~fractions of predicted cases} &\multicolumn{6}{l}{$\Delta \hat{Y}$:~fractions of predicted new cases} \\
%\bottomrule
\hline
\end{tabular}

\caption{\footnotesize Naming convention for the epidemic data, $Y = \{I,R,D\}.$\label{symbols}}

\end{table}

\section{Results}
\subsection{Notation}
Before presenting our findings, we introduce our notation for the epidemic data. Throughout the text, we operate with the infectious ${I}$, recovered  ${R}$, and deceased ${D}$ data. Each dataset is a time series of values, each corresponding to a specific observation date. For brevity, we refer to the triplet of infectious, recovered, and deceased data as $Y = \{I,R,D\}$. All values contained in the $Y$ time series are fractions of individuals found in the corresponding state on a specific day. For instance, $I[k]$ corresponds to the ratio of individuals who are infectious on day $k$ and the population size. Since  COVID-19 reported data often comes in the form of changes in the number of epidemic cases, we find it convenient to introduce the daily changes in epidemic data as $\Delta Y[k+1] \equiv Y[k+1] - Y[k]$ for $Y = \{I,R,D\}$. Further, in this work, we operate with reported epidemic data $\tilde{Y}$ and inferred data $\hat{Y}$. Since reported and inferred data are expected to differ from the true data, we need to distinguish the three. We refer to the true data as $Y$. We summarize our notation in Table~\ref{symbols}.

\subsection{Evidence for reporting delays in epidemiological data}\label{sec_statistical_analysis}

\begin{figure}[!htbp]
\centering
\includegraphics[width=15cm]
{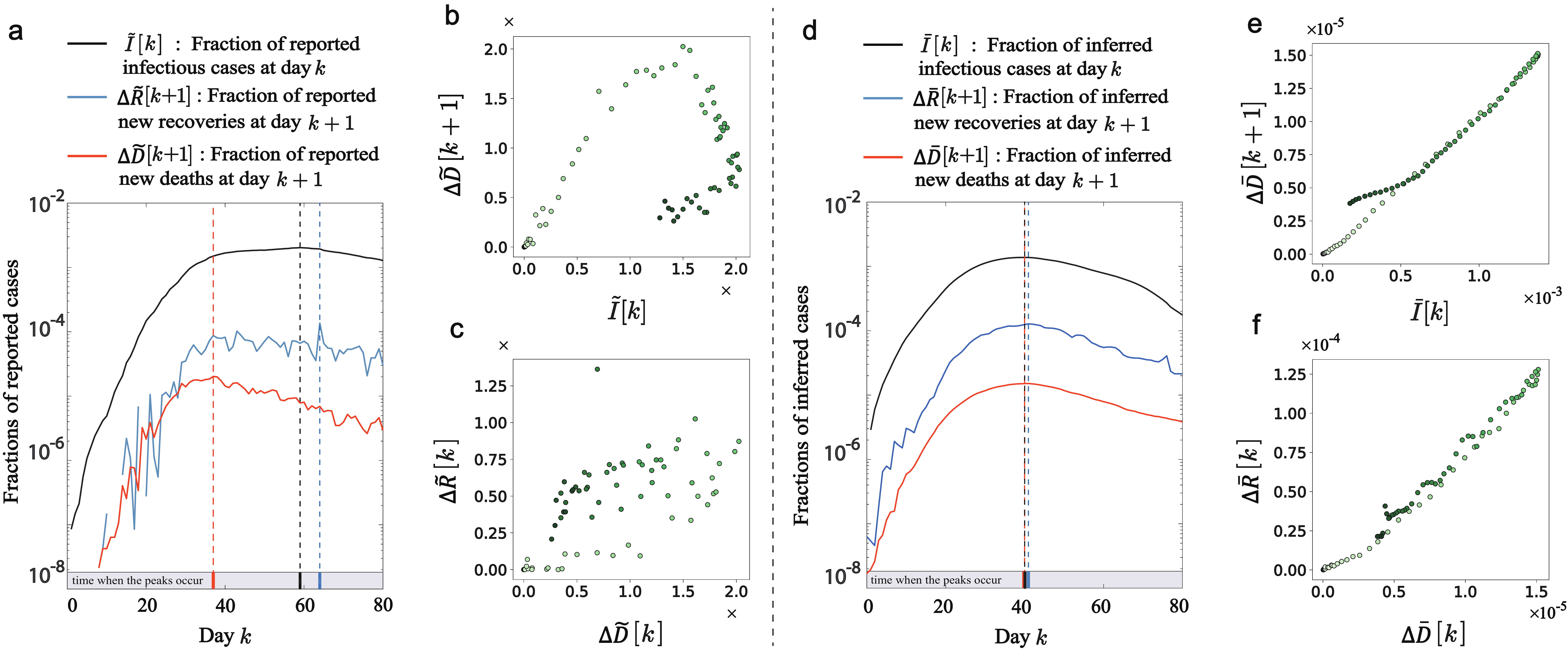}%
\caption{\footnotesize {\bf Evidence for reporting delays in COVID-19 epidemic data.} a, Reported infectious $ \tilde{I}[k] $, recovered  $ \Delta \widetilde{R}[k+1] $ and deceased $ \Delta \widetilde{D}[k+1] $ cases for the first COVID-19 wave in Spain. The $ k=0 $ day corresponds to February 25, 2020. Note the difference in peaks of the reported data. Panels b and c display pairwise color-coded scatter plots of $ \Delta \widetilde{D}[k+1] $ vs $ \tilde{I}[k] $, and $ \Delta \widetilde{R}[k] $  vs $ \Delta \widetilde{D}[k]$ for Spain. Colors, from light to dark green, reflect different days in the data ranging, respectively, from $k=0$ to  $k=80$. Note that the scatter plots in panels b and  c form, respectively, clockwise and counterclockwise loop patterns.  Panels d-f display the epidemic data in Spain after the removal of reporting delays.}%
\label{figure3}%
\end{figure}

We begin the exposition by considering daily reports on the number of infected $\tilde{I}$, recovered  $\tilde{R}$, and deceased $\tilde{D}$ individuals in Spain.  While both the infected and the deceased data indicate that the first COVID-19 wave in Spain peaked in April 2020, the exact timings of the two peaks are, nevertheless, different. As observed in Fig.~\ref{figure3}a, reported new deceased cases $\Delta \widetilde{D}$ reached their peak on April 1st, 2020, while the highest fraction of infectious individuals $\tilde{I}$ was observed 22 days later on April 23, 2020. This observation is not specific to Spain: the peaks in the number of reported new deceased cases precede those of infectious cases by more than one week in most regions, Fig.~S1. We make similar observations for COVID-19 daily recovery reports $\Delta \widetilde{R}$, which exhibit their peaks after those of the deceased cases $\Delta \widetilde{D}$, Fig.~\ref{figure3}a, and Fig.~S1. Furthermore, when plotted as a function of daily deceased cases $\Delta \widetilde{D}$, infectious cases $\tilde{I}$ and daily recovered cases $\Delta \widetilde{R}$ form loop patterns, see Fig.~\ref{figure3}b,c and Fig.~S2.

The patterns observed in Fig.~\ref{figure3}a,b,c, Fig.~S1, and Fig.~S2  may indicate the presence of reporting delays. Indeed, the SARS-CoV-2 virus has hardly changed during the first wave of the pandemic. Therefore,  recovery $\gamma_{r}$ and death $\gamma_{d}$ rates are estimated to be approximately constant during this period~\cite{chen2020clinical,voinsky2020effects,faes2020time}, implying that changes in the fractions of recovered $\Delta \widetilde{R}$, and deceased $\Delta \widetilde{D}$ data are proportional to the fraction of infectious individuals $\widetilde{I}$, contradicting Fig.~\ref{figure3}a,b,c. We hypothesize that the disagreement between $\Delta \widetilde{R}$, $\Delta \widetilde{D}$, and $\widetilde{I}$ is due to reporting delays. If each $\Delta \widetilde{R}$, $\Delta \widetilde{D}$, and $\widetilde{I}$ time series are reported with different delays, their peaks are expected to differ. Likewise, the loop patterns of Fig.~\ref{figure3}b,c  may also be the result of an effective time shift between two non-monotonous time series. The upper part of the loop in Fig.~\ref{figure3}b is due to the fact that initially $\Delta D[k]$  and $I[k]$ both increase as a function of time. The lower part of the loop is observed when $\Delta D[k]$ and $I[k]$ are both decreasing as a function of time. The middle section of the loop pattern corresponds to the time window when $\Delta D[k]$ is decreasing while $I[k]$ is still increasing as a function of time. We verified our hypothesis on synthetic epidemic data with reporting delays, observing similar loop patterns, see Supplementary Information B and Fig.~S3.

\subsection{A null model for reporting delays}\label{sec_statistical_framework}

\begin{figure*}[!htbp]
\centering
\includegraphics[width=9cm]{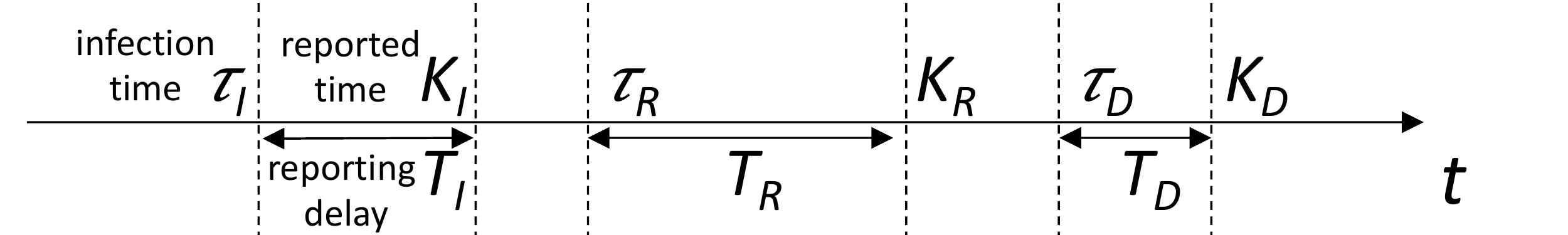}%
\caption{\footnotesize  A schematic representation of epidemic events and corresponding delay times.}%
\label{fig:schematic}%
\end{figure*}

To uncover reporting delays, we employ the following null model.  Within the model, each individual $i$ is endowed with random event time $\tau_{{Y}_i}$ of either getting infected, $\tau_{{I}_i}$, recovered, $\tau_{{R}_i}$, or deceased, $\tau_{{D}_i}$. We denote the random delay time for event $Y_i = \{I_{i}, R_{i}, D_{i}\}$ as $T_{{Y}_i} = \{T_{{I}_i}, T_{{R}_i}, T_{{D}_i} \}$, and the corresponding random reported time as $K_{{Y}_i}=\tau_{Y_i}+T_{{Y}_i}$, Fig.~\ref{fig:schematic}. Assuming that the reporting delays $T_{{Y}_i}$ are independent of events $Y_i$, we obtain for the reported time $K_{{Y}_i}$
\begin{equation}
        \Pr\left[K_{{Y}_i}=k\right]=\sum_{y=0}^{\infty}\sum_{m=0}^{\infty}\Pr\left[\tau_{Y_i}=y\right]\Pr\left[T_{{Y}_i}=m\right]\delta_{y+m,k} =\sum_{m=0}^{k}\Pr\left[\tau_{Y_i}=k-m\right]\Pr\left[T_{Y}=m\right].
  \label{eq:convolution}
\end{equation}
By taking the arithmetic mean of Eq.~[\ref{eq:convolution}], we find that the expected reported fraction of individuals in state $Y$, $\Delta \widetilde{Y}[k] = {\frac{1}{N}}\sum_{i}\Pr\left[K_{Y_{i}}=k\right]$,  is a discrete convolution 
\begin{equation}\label{eqn:real_predict_data}
	\Delta \widetilde{Y}[k]=\sum_{m=0}^{k}\Pr\left[T_{Y}=m\right]\Delta {Y}[k-m],
\end{equation}
where $Y = \{I,R,D\}$. Equation [\ref{eqn:real_predict_data}] serves as the foundation for uncovering reporting delays and improving epidemic forecasts.

\subsection{Statistical framework to uncover reporting delays}\label{sec:estimate_delay}

While the delay inference problem can be formulated and solved in a non-parametric way, to simplify the exposition, we solve the problem parametrically.  We assume that the shapes of delay distributions $\Pr[T_{Y}=m]$ are known, but their parameters $\bm{\kappa}_{Y}$ are not. We then search for $\bm{\kappa}_{Y}$ maximizing the correlations among the epidemic data: $\Delta Y = \{ I, \Delta R,\Delta D\}$.

We consider three families of two-parameter discrete-time distributions:  the negative binomial distribution, the Neyman type A distribution, and the P\'olya-Aeppli distribution, see {\it Materials and Methods A}. These distributions are quite general and contain some well-known 1-parameter distributions as special cases. The logarithmic, geometric, and Poisson distributions are, for instance, all special cases of the negative binomial distribution~\cite{freeman1980fitting}. The three distributions are sufficiently general to generate data with variable mean and variance and of varying skewness~\cite{freeman1980fitting,johnson2005univariate} -- properties expected of the epidemic delay data, see \textit{Materials and Methods} A. For brevity, we focus on the P\'olya-Aeppli distribution in the main text and report the results for the other two distributions in Fig.~S4.

In more precise terms, we assume that the reporting delays correspond to three datasets, $Y = \{ I, R, D\}$, which are all characterized by the P\'olya-Aeppli distribution, albeit with different parameters  $\bm{\kappa}_{Y}\equiv\{\lambda_{Y}, \theta_{Y}\}$. Hence, there are in total 6 parameters for three P\'olya-Aeppli distributions, $ \bm{\kappa}=(\lambda_{I},\theta_{I},\lambda_{R},\theta_{R},\lambda_{D},\theta_{D}) $. 

For prescribed values $\bm{\kappa}$, we can remove the effects of reporting delays by solving  Eq.~[\ref{eqn:real_predict_data}] as a system of linear equations for original epidemic data $\Delta Y$. Since the resulting epidemic data is expected to differ from the true data, we refer to the reconstructed data as $\Delta \bar{Y}_{\bm{\kappa}}$. Given the incremental time series $\Delta \bar{Y}_{\bm{\kappa}}$, the cumulative time series $\bar{Y}_{\bm{\kappa}}$ follows as $\bar{Y}_{\bm{\kappa}}[k]= \bar{Y}_{\bm{\kappa}}[k-1] + \Delta \bar{Y}_{\bm{\kappa}}[k]$ for $k > 1$.

In our framework, we aim to determine parameters $\bar{\bm{\kappa}}$ that maximize the product of pairwise correlations among the three epidemic time series in $Y$:
\begin{equation}
O_{b}(Y) \equiv O_{b}(I,\Delta R, \Delta D) = \rho (\Delta {R}, \Delta {D}) \rho ({I}, \Delta {R}) \rho ({I}, \Delta {D}),
\label{eq:obj_function}
\end{equation}
where $\rho(X,Y)$ is the Pearson correlation coefficient between time series $X$ and $Y$. The objective function $O_{b}( Y)$ in Eq.~[\ref{eq:obj_function}] reaches its maximum value of $1$ when all three pairwise correlations among the $I$, $\Delta R$, and $\Delta D$ time series are 1, which we expect when recovery $\gamma_r$ and deceased $\gamma_d$ epidemic rates are constant. Due to the nature of the Pearson correlation coefficient, the objective function $O_{b}(Y)$ is invariant under the constant time shift $T$ of the epidemic data,  $O_{b}({Y}[k]) = O_{b}({Y}[k-T])$. As a result, we can only infer the reporting delays up to a constant time shift $T$.

To maximize $O_{b}(Y)$, we use the random search~\cite{bergstra2012random}: we conduct a large set of independent random iterations, $\ell = 1,\ldots,L$. At each iteration $\ell$, we select the elements of the parameter vector $ \bm{\bar{\kappa}}$ uniformly at random from the prescribed domain of values. Further, at each iteration  $\ell$, we use the selected parameter set $\bm{\bar{\kappa}}_{\ell}$ to reconstruct the original epidemic data $\bar{Y}_{\bm{\bar{\kappa}}_{\ell}}$ by solving Eq.~[\ref{eqn:real_predict_data}].  We then compute the pairwise correlations in $\bar{Y}_{\bm{\bar{\kappa}}_{\ell}}$ to obtain the objective function $O_{b}\left(\bar{Y}_{\bm{\bar{\kappa}}_{\ell}}\right)$. After completing all random search iterations, the resulting delay parameter $\bm{\hat{\kappa}}$ is the one maximizing the objective function, $\bm{\hat{\kappa}} = {\rm argmax}_{\bm{\bar{\kappa}}_{\ell}} O_{b}\left(\bar{Y}_{\bm{\bar{\kappa}}_{\ell}}\right)$, see {\it Materials and Methods B}. 

\subsection{Tests on synthetic data}\label{sec:synthetic_data}
Before uncovering reporting delays in real epidemic data, we test our framework on synthetic datasets, which we generate with the SIRD compartmental model. Within the SIRD model, the population is split into four compartments: susceptible $ S $, infectious $ I $, recovered $R$, and deceased $D$. Compartment $ S $ denotes the fraction of susceptible individuals who can be infected by infectious individuals. Compartment $I$ denotes the fraction of individuals who have been infected but have not recovered or are deceased. Compartments $R$ and $D$ are respectively the fractions of individuals who have recovered or are deceased. The SIRD model assumes that recovered individuals become immune and cannot be infected by the virus in the future. Discrete-time transitions between the compartments are governed by first-order difference time equations
\begin{equation}
\begin{aligned}
&I[k+1]-I[k]  = \beta {I[k]S[k]}-(\gamma_r+\gamma_d){I[k]},\\
&\Delta R  [k+1]\equiv R[k+1]-R[k] = \gamma_{r} {I[k]},\\
&\Delta D  [k+1] \equiv D[k+1]-D[k] = \gamma_{d} {I[k]},\\
&\Delta I  [k+1] \equiv I[k+1]-I[k]+\Delta R  [k+1]+\Delta D  [k+1],\\
&{S[k]} + {I[k]} + {R[k]} + {D[k]} = 1,
\end{aligned}
\label{eqn:discrete_SIRD_model}
\end{equation}
where $\beta$, $\gamma_r$, and $\gamma_d$ are model parameters quantifying, respectively, the infection rate, the recovery rate, and the deceased rate. 

To test our framework, we generated $50$ SIRD epidemic model datasets with different epidemic parameters and added synthetic reported delays to the obtained times series, as prescribed by Eq.~[\ref{eqn:real_predict_data}]. After generating the synthetic data we  forgot the parameters of reporting delays and used our statistical framework to infer them.  Figures~\ref{real_rep_infer}a,b indicate that the inferred delay parameters are in good agreement with the true parameters that generated the datasets. We observe that the inference errors decrease fast as a function of the number of iteration steps saturating at the margin of 2 days after $10^{7}$ iterations, Fig.~\ref{real_rep_infer}a. To assess the robustness of the inference procedure, we have conducted cross-inference experiments by generating synthetic data with one delay distribution and inferring delays using another distribution, arriving at similar results, Fig.~S4.

\subsection{Uncovering reporting delays in real data}\label{sec:real_data}
After testing our inference framework on synthetic data, we moved on to uncover reporting delays in real epidemic data that we have collected from 8 regions worldwide, Supplementary Information A. Figs.~\ref{figure3}d-f display the reconstructed epidemic data for Spain after the identification and removal of reporting delays, see Table~S2 for the inferred parameters.

Consistent with our expectations, the reconstructed infected, recovered, and deceased time series are strongly correlated, Figs.~\ref{figure3}e,f, and their peaks co-occur at day 40, Fig.~\ref{figure3}d. By comparing the original correlation coefficients  $O_{b}(\widetilde{Y})$ with their optimized counterparts $O_{b}(\bar{Y})$ we discover that the latter are significantly increased with a relative improvement ranging from $0.94$ to $9.5$ after the reconstruction, Table~S2.

We summarize the properties of uncovered reporting delays in the 8 regions in Table~2. We find that the infectious and recovered data delays are longer than those for deceased cases and vary from several days to several weeks. This observation is hardly surprising. Indeed, there are several factors contributing to the delays of infectious cases. One factor is the delay between an individual becoming infectious, and the symptom onset~ \cite{sun2021transmission}. Another factor, particularly significant in the first COVID-19 wave, is the delay between the symptom onset and the test result~\cite{kretzschmar2020impact}. In their turn, delays in the recovered events are likely caused by hospital discharge policies. This is the case since the recovered data is usually derived from hospital discharge events, which occur after the patient's recovery.

Based on the expected delay values, one can naturally split the 8 ROIs into three categories: (i) large infectious and small recovered delays: Romania, Germany, and Denmark, (ii) small infectious and large recovered delays: Italy and Spain, and (iii) large infectious and recovered delays: Wuhan, Hubei, and Turkey. 

Small infectious delays imply that COVID-19 tests are timely and accurate. This seems to be the case for Italy, which executed more tests per capita in April 2020 than other countries \cite{WinNT}. On the contrary, the testing ability of the Hubei province was significantly insufficient during the first wave of the COVID-19 outbreak.

Large delays in the recovered data may be attributed to strict hospital discharge policies. Discharge policy in Italy, for instance, was based on the negative test~\cite{jespers2020international}, and it has been shown that COVID-19 tests may stay positive for an extended time after COVID-19 symptoms disappear.  In contrast, discharge policies in Denmark were based not on the negative test but on patient symptoms~\cite{jespers2020international}, likely leading to shorter reporting delays in the recovered data.

Large standard deviations in the reporting delays may indicate irregularities in reporting mechanisms. As an example, Hubei province expanded its daily testing capacity from 200 to 2,000 individuals from the beginning of the pandemic until January 27th \cite{WinNT2}. As a result, fewer individuals were tested late at the end of the first COVID-19 wave compared to its beginning, likely resulting in the large standard deviation of infectious data delays.

\begin{figure}[!htbp]
\centering
\includegraphics[width=15cm]{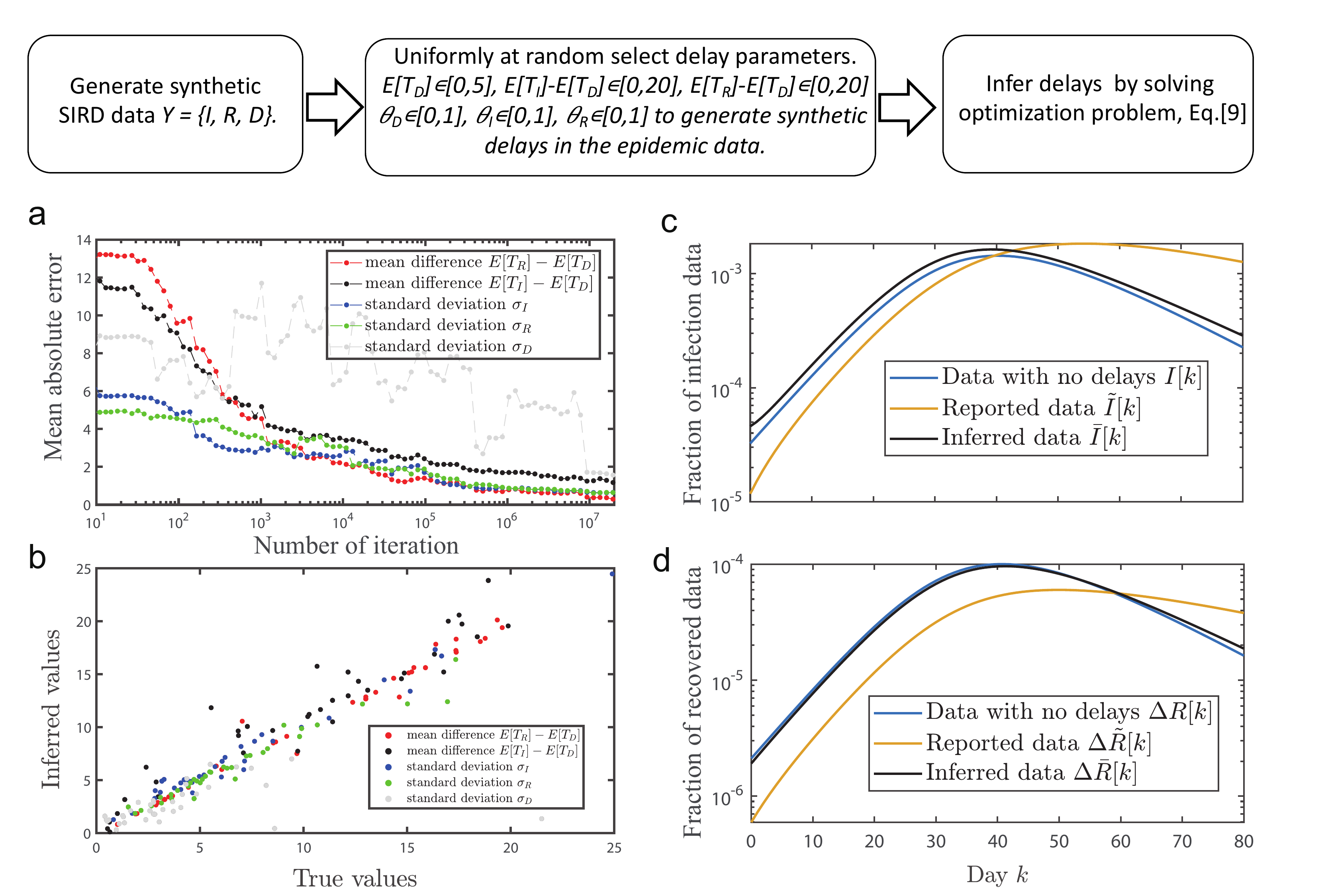}%
\caption{  
{\bf Uncovering reporting delays in synthetic data.} We generate epidemic data using the SIRD model, Eq.~[\ref{eqn:discrete_SIRD_model}] with parameters $\beta = 0.23$, $\gamma_r = 0.079$, and $\gamma_{d} = 0.11$ and variable delay parameters. The schematics on top illustrates the process of generation and inference on synthetic data.  a, The mean absolute inference errors as functions of the number of iteration steps in the random search. Due to the nature of the Pearson correlation coefficient, it is only possible to infer relative delays, $E\left[T_{R}\right] - E\left[T_{D}\right]$, $E\left[T_{I}\right] - E\left[T_{D}\right]$, and $E\left[T_{R}\right] - E\left[T_{I}\right]$.  b, The relationship between the true and the inferred values of delay parameters. (c,d) An example of an SIRD synthetic data (c) before and (d) after the removal of synthetic reporting delays.}
\label{real_rep_infer}%
\end{figure}

The small standard deviation in the delays of recovered data observed in Hubei and its capital Wuhan is likely the consequence of the strict discharge criteria~\cite{fu2021database}. Although strict discharge policies cause significant delays, these delays are similar, resulting in relatively smaller $\sigma_R$ values. In contrast to Chinese regions, the recovery data for Germany are not reported directly but instead are estimated by a not explained algorithm \cite{WinNT3}, resulting in large errors and, consequently, larger $\sigma_R$ values. Based on the optimized value $O_{b}(\bar{Y})$ as shown in Table~\ref{real_rep_infer}, the optimization performances of our algorithm for Italy, Spain, and Turkey are better than the other countries. 

\begin{table}
	\centering
	\begin{tabular}{cccccccccc}
		\hline
		Regions& $E[T_R]-E[T_D]$& $E[T_I]-E[T_D]$ & $\sigma_I$& $\sigma_R$ & $\sigma_D$ & $\rho(\Delta \bar{R},\Delta \bar{D})$ & $\rho(\bar{I},\Delta \bar{R})$ & $\rho(\bar{I},\Delta \bar{D})$ & $O_{b}(\bar{Y})$ \\
		\hline
		Italy & 28.52 & 6.59 & 5.08& 27.79 & 0.72 & 0.91 & 0.94 & 0.99 & 0.84\\
		Spain & 22.66 & 6.36 & 8.63 & 28.39 & 0.73 & 0.99 & 0.99 & 1.00 & 0.98\\ 
		Wuhan & 14.80 & 20.64 & 51.67 & 16.13 & 23.87 & 0.78 & 0.89 & 0.91 & 0.63\\
		Turkey & 14.13 & 24.85 & 43.89 & 12.47 & 11.21 & 0.91 & 0.95 & 0.98 & 0.85\\  
		Hubei & 9.96 & 21.07 & 78.62 & 10.89 & 54.14 & 0.85 & 0.90 & 0.92 & 0.71\\
		Romania & 3.30 & 19.05 & 43.08 & 90.12 & 0.29 & 0.81 & 0.89 & 0.97 & 0.70\\
		Germany & 2.72 & 16.55 & 39.25 & 106.89 & 4.31 & 0.87 & 0.88 & 0.99 & 0.76\\
		Denmark & 0.17 & 25.07 & 36.90 & 2.71 & 28.25 & 0.83 & 0.93 & 0.94 & 0.72\\  
		\hline
	\end{tabular}
	\caption{\footnotesize Inferred reporting delays. The table displays the inferred differences between the expected delay times $E[T_R]-E[T_D]$, $E[T_I]-E[T_D]$,  standard deviations $\sigma_I$, $\sigma_R$ and $\sigma_D$, and optimized values for the objective function $O_{b}(\bar{Y})$. See Tables~S1 and S2 for the corresponding parameters of the P\'{o}lya-Aeppli distributions.} \label{time_delays}
\end{table}

\begin{figure}[htbp]
\centering
\includegraphics[width=13cm]
{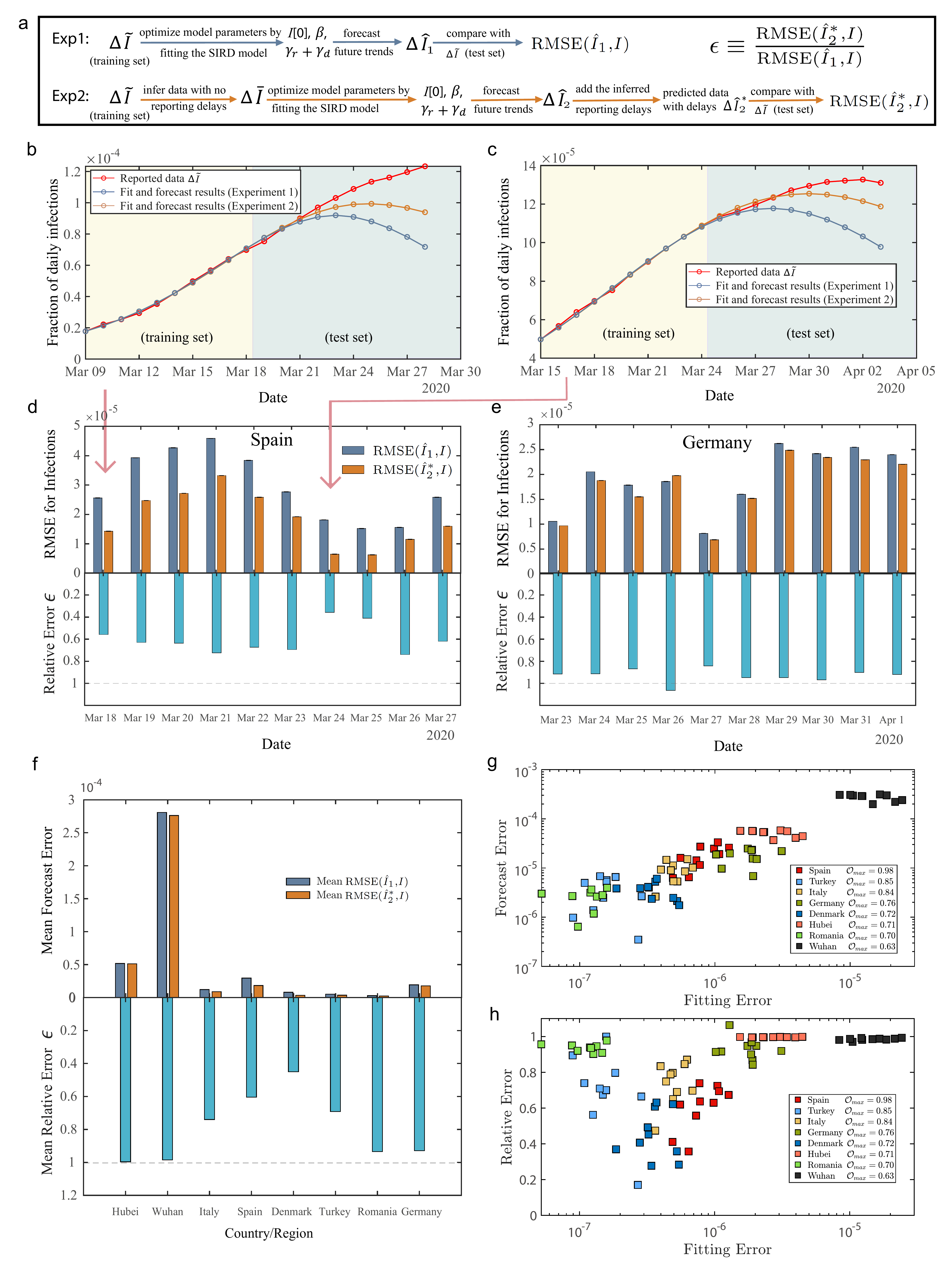}%
\caption{\footnotesize {\bf Accounting for reporting delays improves epidemic forecasts} a, The schematic diagram for the two forecast experiments. b, c display the results of epidemic forecasts in Spain with different forecast start dates. The forecasts of experiment 2 (yellow) are closer to the reported data  (red) than the forecasts of experiment 1(blue).  The forecast results for all 8 countries or regions are shown in Figs.~S13-S14. d,e, All forecast results for Spain and Germany. Shown are the RMSE forecast errors and their ratios. h, Average forecast errors $\epsilon$ for the 8 regions. The benefits of removing reporting delays vary by region. g, The average RMSE forecast error of experiment 2 as a function of SIRD model fitting error for all experiments. h, The mean relative forecast error as a function of SIRD model fitting error for all experiments.}%
\label{figure6}%
\end{figure}

\subsection{Improving epidemic forecasts}
Is accounting for reporting delays likely to improve epidemic forecasts? To answer this question, we designed two experiments. Experiment~$1$ aims to forecast the epidemics ignoring reporting delays and serves as a baseline for Experiment~$2$, which uncovers reporting delays prior to forecasting epidemic data. 

In both experiments, we split the reported epidemic data into two parts, which we call the training and the testing sets, respectively, see Fig.~\ref{figure6}a. In Experiment 1, we fit the training set with the SIRD epidemic model, obtaining model parameters $\beta$, $\gamma_r+\gamma_d$, and the fraction of initial infected cases $I[0]$. We then use the SIRD model with the obtained parameters to forecast the epidemic data, Supplementary Information C.

In Experiment 2, we first use the reported data $ \widetilde{Y}$ in the training set to infer the parameters of the delay distributions $\bar{\bm{\kappa}}$. We rely on these parameters to remove reporting delays from the testing set and obtain reconstructed data $\Delta \bar{I}$. In the next step, we fit the reconstructed  data $\Delta \bar{I}$ with the SIRD model, obtaining $\beta$, $\gamma_r+\gamma_d $ and $I[0]$ spreading parameters. We use these spreading parameters to forecast the epidemic data $\Delta \hat{I}$, which we compare to the reported data $\Delta \widetilde{I}$ in the testing set. Since $\Delta \widetilde{I}$ in the testing set contains the reporting delays, while the forecast $\Delta \hat{I}$ does not,  we added the inferred reporting delays back to the $\Delta \hat{I}$ using Eq.~[\ref{eqn:real_predict_data}], obtaining $\Delta \hat{I}^*$, See Fig.~\ref{figure6}a and Supplementary Information C.

Figures~\ref{figure6}b,c present the results of the forecast experiments for Spain, indicating that correcting for reporting delays does improve the forecast accuracy.  To evaluate the forecast errors in  a systematic way, we evaluated the root mean square errors between the forecasts $I_F$ and the testing $I_T$ sets:
\begin{equation}
 {\rm RMSE}(I_F,I_T)  = \left[\frac{1}{n} \sum_{k=0}^{n-1}  (I_{F}[k] - I_{T}[k] )^2 \right]^{\frac{1}{2}},
\end{equation}
where $ n $ is the size of the testing set. Further, to quantify the benefits of accounting for reporting delays, we used the ratio of the root mean square errors measured in forecasts with and without reporting delays, $\epsilon \equiv \frac{{\rm RMSE}(\hat{I}^{*}_{2},I) }{{\rm RMSE}(\hat{I}_{1},I) }$, where $\hat{I}_{1}$ and  $\hat{I}^{*}_{2}$ are the epidemic forecasts obtained in experiments 1 and 2, respectively. The smaller the ratio, the smaller the relative forecast error.

We measured the accuracy of epidemic forecasts with different start dates for Spain in Germany. While we did observe a substantial improvement in forecast accuracy for Spain, Fig.~\ref{figure6}d, this was not the case for Germany, where the removal of reporting delays only improved the forecasts by a small margin, Fig.~\ref{figure6}e. On a broader scale, we observed that the benefits of correction for reporting delays vary across all regions of interest, Fig.~\ref{figure6}f, Fig.~S5, and Fig.~S6. While there is nearly a two-fold improvement in the forecast accuracy for Denmark, there is little improvement for Wuhan and Hubei. 

There are two factors that may hinder epidemic forecast accuracy. The first factor is the insufficient accuracy of reporting delay removal. The second one is the inability of the SIRD model to reproduce the COVID-19 dynamics accurately. We can quantify the former by the maximum attained value of the objective function $O_{b}(\bar Y)_{\rm max}$ that we aim to maximize when removing reporting delays. While there is no direct way to quantify the goodness of the SIRD model in epidemic forecasts, as an indirect measure we use the SIRD fitting error obtained by fitting the training set with the SIRD model. As seen in 
Fig.~\ref{figure6}g, the forecast error is strongly correlated with the SIRD fitting error, Pearson $r = 0.89$, $p < 10^{-27}$, indicating that the highest epidemic forecast accuracy is attained when the model fit is accurate. When it comes to the benefits of removing reporting delays, we observe that the forecast error ratio depends both on the $O_{b}(\bar Y)_{\rm max}$ and SIRD model fitting, Fig.~\ref{figure6}h. We observe the largest error ratio $\epsilon$ in Wuhan, Hubei, and Romania, Fig.~\ref{figure6}h. Wuhan and Hubei  correspond to the largest SIRD fitting errors, while Romania corresponds to the lowest SIRD fitting errors. At the same time, all three regions are quantified by the lowest $O_{b}(\bar Y)_{\rm max}$ values. The other 5 regions, corresponding to smaller error ratios, are characterized by significantly larger $O_{b}(\bar Y)_{\rm max}$ values, Fig.~\ref{figure6}g and Table~2.

These observations indicate that $O_{b}(\bar Y)_{\rm max}$ may be used as an early indicator of the benefit of delay removals. Indeed, lower $O_{b}(\bar Y)_{\rm max}$ values may indicate that reporting delays were not removed successfully or the initial assumption of the proportionality among $I$, $\Delta R$, and $\Delta D$ times series does not hold. We note that $O_{b}(\bar Y)_{\rm max}$ is a good indicator regardless of the SIRD fitting error. In the case of Romania, SIRD fitting errors are among the smallest resulting in low forecast errors. Yet, the removal of reporting delays does not result in even lower forecast errors. 

There might be multiple reasons for suboptimal reporting delay removals in the case of Hubei, Wuhan, and Romania. One possibility is the non-stationary nature of delays. The main assumption of our framework is that delay times are independent and drawn from the same distribution. While COVID-19 did not significantly mutate over the short time span of the first wave, our ability to handle the infections improved significantly. The PCR test capacity in China has grown remarkably during the first COVID wave, possibly explaining the limited effect of reporting delays on improving epidemic forecasts in Hubei and Wuhan.

\section{Discussion}
Data delays are ubiquitous in data sciences and adversely affect data analysis~\cite{akbarov2013warranty}. The unique property of the epidemic data, enabling us to identify and filter out reporting delays, is the proportionality between the number of infectious individuals $I$ and the rates of change in the deceased $\Delta D$ and the recovered $\Delta R$ individuals. 

We relied on this proportionality property to develop a parametric statistical framework to uncover reporting delays in the first COVID-19 wave and applied it to 8 regions significantly affected during the first COVID-19 wave. The character of uncovered delays varies across studied regions and can be explained by region-specific medical capacities and epidemic policies.

Concerning the epidemic forecasts, we found that the benefits of  curated epidemic data, as opposed to the raw data, are maximized in situations when reporting delays are removed efficiently. One of the main factors hindering the efficiency of delay removal -- the non-stationarity of reporting delays -- can be caused by either varying spreading properties of the virus or by rapid changes in medical capacities or epidemic restrictions across the regions. While the spreading properties of COVID-19  did not change significantly during the first wave of the pandemic, both medical capacities and epidemic policies were the subjects of constant updates as the society learned how to handle the pandemic. 

In conclusion,  our statistical framework does not make strong assumptions about the spreading mechanisms of a pathogen.  Be it a well-studied viral infection or a novel virus, our framework should be instrumental in uncovering reporting delays, as long as the spreading properties of the virus remain unchanged during the observation period. Our framework can be used as a preliminary filter for any epidemic forecast tool that takes infected, recovered, and deceased data as input. Accurate and timely epidemic forecasts are of immense value for society and policymakers to minimize the adverse effects of the virus.

\section{Methods}
\subsection{Types of distributions}\label{discrete_distribution}
To determine the family of reporting delay distributions that best suit our data, we consider three different two-parameter discrete distributions \cite{ord1972families} below:\\
(\uppercase\expandafter{\romannumeral1}) Negative binomial distribution.
\begin{equation}
\Pr[T=m]=\binom{m+r-1}{m} (1-p)^m p^r.
\label{eqn:pdf_Negative}
\end{equation}
The negative binomial distribution with parameters $r > 0 $ and $ p \in [0,1]$ has mean value $ E[T]=r(1-p)/p $ and variance $ Var[T]=r(1-p)/p^2 $.\\
(\uppercase\expandafter{\romannumeral2}) P\'{o}lya-Aeppli distribution is also known as the geometric Poisson distribution. \begin{equation}
\Pr[T=m]=\begin{cases} \sum_{j=1}^m e^{-\lambda}\frac{\lambda^j}{j!}(1-\theta)^{m-j}\theta^j\binom{m-1}{j-1}, & m>0 \\ e^{-\lambda}, & m=0 \end{cases}.
\label{eqn:pdf_Polya}
\end{equation}
The P\'{o}lya-Aeppli distribution with parameters $ \lambda>0 $ and $ \theta\in [0,1]$ has mean value $E[T]=\lambda/\theta $ and variance $ Var[T]=\lambda (2-\theta)/\theta^2 $.\\
(\uppercase\expandafter{\romannumeral3}) Neyman type A distribution. 
\begin{equation}
\Pr[T=m]=\dfrac{\mu^m e^{-\xi}}{m!}\sum_{j=0}^{\infty}\dfrac{{(\xi e^{-\mu})}^j}{j!}j^m.
\label{eqn:pdf_Neyman}
\end{equation}
The Neyman type A distribution with parameters $ \xi>0 $ and $ \mu>0$ has mean value $ E[T]=\xi \mu $ and variance $ Var[T]=\xi\mu(1+\mu) $.

\subsection{Inferring reporting delays}\label{infer_approach}
To uncover reporting delays in epidemic data, we determine parameters of delays distributions $\bm{\kappa}_{Y}$ maximizing the pairwise correlations between the
epidemic time series, Eq.~[\ref{eq:obj_function}]. This optimization problem can be compactly written as:
\begin{align}
\begin{aligned}
\label{eq:optimization} 
& \underset{\bm{\kappa}}{\operatorname{arg~max}} \quad O_{b}(Y) \equiv \rho (\Delta R, \Delta D) \rho (I, \Delta R) \rho (I, \Delta D)\\
\text{s.t.}  ~& ~~~ \Delta {I}=\Psi^{-1}_{I} \Delta \widetilde{I},~ \Delta R =\Psi^{-1}_{R} \Delta \widetilde{R},  ~\Delta D =\Psi^{-1}_{D} \Delta \widetilde{D},\\
&\quad \min(\Delta {I}[k], \Delta {R}[k], \Delta {D}[k], {I}[k])\geq 0, \quad\text{for}~~ k=1,\ldots,T.
\end{aligned}
 \end{align}
Here $T$ is the size of epidemic time series, $\widetilde{Y} \equiv \{\widetilde{I}, \widetilde{R}, \widetilde{D}\}$ and ${Y} \equiv \{{I}, {R}, {D}\}$ are reported and reconstructed epidemic data, respectively, while $\Delta Y =\Psi^{-1}_{Y} \Delta \widetilde{Y}$ are the matrix solutions of Eqs.~[\ref{eqn:real_predict_data}]. Indeed, Eqs.~[\ref{eqn:real_predict_data}] can be written in the matrix form as $\Delta  \widetilde{Y} =\Psi_{Y} \Delta{Y}$ for $Y = \{I,R,D\}$, where
\begin{align*}
&{\Psi_{Y}}_{i, j}
\triangleq
\begin{cases}
{\Pr}[T_{Y}=i-j] & \textrm{if} \quad i\geq j. \\
0 & \textrm{otherwise}.
\end{cases},
\end{align*}
Then $\Psi^{-1}$ is the inverse of $\Psi_{Y}$ and $\Delta  {Y} =\Psi^{-1}_{Y} \Delta \widetilde{Y}$.

In the main text, we  assume that reporting delays are iid random variables drawn from three P\'{o}lya-Aeppli distributions,  Eq.~[\ref{eqn:pdf_Polya}], with distinct parameters $\{\lambda_I, \theta_{I}\}$, $\{\lambda_R, \theta_{R}\}$, and $\{\lambda_D, \theta_{D}\}$. In the case of P\'{o}lya-Aeppli distributions,  the parameter vector takes the form of $\bm{\kappa} \equiv \{\lambda_I, \theta_{I}, \lambda_R, \theta_{R}, \lambda_D, \theta_{D}\}$. We solve the optimization problem given by Eq.~[\ref{eq:optimization}] and the random search~\cite{bergstra2012random}. For each iteration $\ell = 1,\ldots, L$ we treat the expected values of the P\'{o}lya-Aeppli distributions, $E\left[T_{Y}\right] = \lambda_Y/\theta_Y$, as iid random variables and draw from the uniform pdfs $U[0,30]$ for $Y = \{I,R,D\}$. Similarly, we draw 
$\theta_Y$ parameters independently at random from uniform pdfs $U[0,1]$ and then determine $\lambda_{Y}$ parameters as $\lambda_{Y} = \theta_{Y} E\left[T_{Y}\right]$, obtaining $\bm{\kappa}_{\ell}$. In our experiments we set the maximum number of search iterations to $L = 10^{7}$.

For each $\bm{\kappa}_{\ell}$ we use Eq.~[\ref{eqn:real_predict_data}] to reconstruct original data $Y_{\ell}$, which we then use to compute the objective function $O_{b}\left(Y_{\ell}\right)$. After completing all iterations, the thought parameter vector $\bm{\hat{\kappa}}$ describing reporting delays is the one corresponding to the maximum  $O_{b}\left(Y_{\ell}\right)$ value, 
$\bm{\hat{\kappa}} = \operatorname{arg~max}_{\bm{\kappa}} \quad O_{b}(Y)$.

\section*{Data availability}
All epidemic data used in this work are publicly available from the original sources, see Supplementary Information A.  The extracted epidemic time series for the eight regions of interest are deposited in FigShare (\url{https://doi.org/10.6084/m9.figshare.22639519.v1}).

\section*{Acknowledgements} This research has been funded by the European
Research Council (ERC) under the European Union's Horizon 2020 research and innovation programme (grant agreement No 101019718). L. M. is thankful for the support from the China Scholarship Council. M. K. has been supported by the Dutch Research Council (NWO) grant OCENW.M20.244.

%\bibliographystyle{naturemag}
%\bibliography{lm_mk_bib}

\end{document}